\documentclass[prd,aps,twocolumn
,preprintnumbers,nofootinbib,showpacs,amsmath,amssymb,superscriptaddress]{revtex4-1}
\usepackage[dvipdfmx]{graphicx}
\usepackage{amsmath, amssymb}
\usepackage{comment}
\usepackage{color}
\usepackage{float}
\usepackage{braket}
\usepackage{color} 
\usepackage{amsmath,braket}
\usepackage{amssymb}
\usepackage{bm}
\usepackage[dvipdfmx]{graphicx} 
\newcommand{\bea}{\begin{eqnarray}}
\newcommand{\eea}{\end{eqnarray}}
\newcommand{\be}{\begin{equation}}
\newcommand{\ee}{\end{equation}}

 \makeatletter  
\def\alt{\mathrel{\mathpalette\gl@align<}}
\def\agt{\mathrel{\mathpalette\gl@align>}}
\def\gl@align#1#2{ \lower.6ex\vbox{\baselineskip\z@skip\lineskip\z@
\ialign{ $\m@th#1\hfil##\hfil$\crcr#2\crcr\sim\crcr }} } \makeatother

\makeatletter
\@addtoreset{equation}{section}

\makeatother

\makeatletter
\newcommand{\cdotfill}[1]{%
\leavevmode
\cleaders \hb@xt@#1{\hss$\cdot\m@th$\hss }\hfill \kern \z@}
\begin{document}

\preprint{KEK-TH-1942,RESCEU-31/16,HUPD-1609}
\vspace{2cm}
\title{Quantum radiation produced by the entanglement of quantum fields}
\vspace{1cm}

\author{Satoshi Iso}
\affiliation{KEK Theory Center, High Energy Accelerator Research Organization (KEK)}

\author{Naritaka Oshita}
\affiliation{Department of Physics, Graduate School of Science, The University of Tokyo, Bunkyo-ku, 
  Tokyo 113-0033, Japan }
\affiliation{Research Center for the Early Universe (RESCEU),Graduate School of Science, 
The University of Tokyo, Bunkyo-ku, Tokyo 113-0033, Japan}

\author{Rumi Tatsukawa}
\affiliation{Graduate school of Physical Sciences, Department of Physical Sciences,
Hiroshima University, Higashi-hiroshima, Kagamiyama 1-3-1, 739-8526, Japan}

\author{Kazuhiro Yamamoto}
\affiliation{Graduate school of Physical Sciences, Department of Physical Sciences,
Hiroshima University, Higashi-hiroshima, Kagamiyama 1-3-1, 739-8526, Japan}

\author{Sen Zhang}
\affiliation{Okayama Institute for Quantum Physics, Kyoyama 1-9-1, Kita-ku, Okayama 700-0015, Japan}

\begin{abstract} 
We investigate the quantum radiation produced
by an Unruh-De Witt detector in a uniformly accelerating motion coupled to the
vacuum fluctuations. Quantum radiation is nonvanishing, which is consistent with  the previous calculation by Lin and Hu 
[Phys.~Rev.~D~{\bf73}, 124018 (2006)].
We infer that this quantum radiation from the Unruh-De Witt detector is generated 
by the nonlocal correlation of the Minkowski vacuum state, which has its origin in
the entanglement of the state  between the left and the right  Rindler wedges. 
\end{abstract} 

\maketitle

%\begin{multicols}{2}

\section{Introduction}
An accelerated observer 
sees the Minkowski vacuum state as a thermally excited state, which is 
characterized by the Unruh temperature $T_U=a/2\pi$, where $a$ is the acceleration. 
By the equivalence principle \cite{Unruh,UnruhWald},
the Unruh effect can be understood in analogy with the Hawking radiation, 
which predicts the thermal radiation from black holes.
Since  both relativity and  quantum mechanics simultaneously play important roles 
in these effects,  detection of the Unruh effect  will have a big impact on the research 
of fundamental physics (cf. \cite{SokolovTernov}).

Signals of the Unruh effect will be tiny since the  Unruh 
temperature is very low, $T_U=4\times 10^{-20}(a/9.8[{\rm m/s^2}])$K for typical values of acceleration. 
Chen and Tajima pointed out a nice idea of testing the Unruh effect 
using intense laser's electric field for accelerating an electron,
which has inspired many  following works \cite{ChenTajima,Schutzhold,Schutzhold2,ELI}.
However,  subsequent investigations demonstrated that naively expected quantum 
radiations from thermal random motions induced by the Unruh effect almost
cancel out due to the interference effect \cite{IYZ,OYZ15,OYZ16}.
These works also showed the cancellation is not complete 
and some quantum radiation remains, though 
 its physical origin  is not well understood.

In order to clarify the possible signature of the Unruh effect in
the quantum radiation, we revisit the problem of the quantum
radiation emanated from an Unruh-De Witt detector in the uniformly
accelerating motion \cite{Raine,Raval,LH,Lin,IYZ2013}.  
We find nonvanishing quantum radiation, which is consistent with 
the previous calculation by Lin and Hu \cite{LH}.
We point out that this quantum radiation is related to the nonlocal correlation 
nature of the Minkowski vacuum state, which has its origin in
the entanglement of the state  between the left and the right  Rindler wedges. 

This paper is organized as follows. In section 2, we review the model 
of the  Unruh-De Witt detector coupled to a massless scalar field. 
In section 3, we derive the nonvanishing quantum radiation form the 
the  Unruh-De Witt detector. In section 4, we discuss about the origin 
of the nonvanishing quantum radiation. Section 5 is devoted to 
summary and conclusions. In the appendix, a mathematical formula 
to describe the quantum radiation flux is presented. 

%%%%%%%%%%%%%%%%%%%%%%%%%%%%%%%%%%%%%%%%%%%%%%%%%%%%%%%%%%%%%%%%%%%
\section{Unruh-De Witt detector model}
%%%%%%%%%%%%%%%%%%%%%%%%%%%%%%%%%%%%%%%%%%%%%%%%%%%%%%%%%%%%%%%%%%%
We consider the model consisting of a massless
scalar field $\phi$ and a harmonic oscillator $Q$,
which we call an Unruh-De Witt detector, described by  the action,
\begin{eqnarray}  
&&S[Q,\phi ; z] =
\frac{m}{2} \int d \tau \left( \dot{Q}^2(\tau) - \Omega_0^2 Q^2(\tau) \right) 
\nonumber \\
&&~~+ {1\over 2}\int d^4 x \partial^\mu \phi(x) \partial_\mu \phi(x)  
\nonumber \\
&&~~
+ \lambda \int d^4 x d\tau Q(\tau) \phi(x) \delta^{(4)}_D(x-z(\tau)),
\end{eqnarray} 
where $m$ and $\Omega_0$ are the mass and the angular frequency 
of the harmonic oscillator, respectively, $\lambda$ is the coupling constant,
and $\delta^{(4)}_D(x-y)$ is the 4-dimensional Dirac delta function. 
The world line trajectory of the detector is specified by
$x^\mu=z^\mu(\tau)$, where $\tau$ is the proper time of the detector.
We consider the trajectory in a uniformly accelerated motion 
$z^\mu(\tau)=a^{-1}\left(\sinh{a\tau},\cosh{a\tau},0,0\right)$.
%where $a$ is the acceleration.
Equations of motion for $Q(\tau)$ and $\phi(x)$ are given by 
\begin{eqnarray}
&&\ddot Q(\tau)+\Omega_0^2Q(\tau)={\lambda\over m}\phi(z(\tau))
\label{eqQ}, \\
&&\partial^2\phi(x)=\lambda\int d\tau Q(\tau)\delta^{(4)}_D(x-z(\tau)).
\label{eqphiA}
\end{eqnarray}
The solution of the scalar field is written as a sum of the homogeneous 
solution $\phi_{\rm h}(x)$ and the inhomogeneous solution $\phi_{\rm inh}(x)$, i.e.,
$\phi(x)=\phi_{\rm h}(x)+\phi_{\rm inh}(x)$. $\phi_{\rm inh}(x)$ is given by
$\phi_{\rm inh}(x)=\lambda\int d\tau Q(\tau)G_R(x-z(\tau))$,
where $G_R(x-y)$ is the retarded Green function of the massless scalar field. 
Using the regularized retarded Green function, (\ref{eqQ}) becomes
\begin{eqnarray}
%\left({d\over d\tau^2}+2\gamma{d\over d\tau}+\Omega^2\right)Q(\tau)
\ddot Q(\tau)+2\gamma\dot Q(\tau)+\Omega^2Q(\tau)
={\lambda\over m}\phi_{\rm h}(z(\tau)),
\label{eqQ2A}
\end{eqnarray}
where we introduced $\gamma=\lambda^2/8\pi m$ and the renormalized frequency $\Omega$
(see Ref.~\cite{LH}).

Using the Fourier transformations,
\begin{eqnarray}
&&Q(\tau)={1\over2\pi}\int_{-\infty}^{\infty} d\omega e^{-i\omega\tau}\tilde Q(\omega),
\label{solQtau}
\\
&&\phi_{\rm h}(z(\tau))={1\over2\pi}\int_{-\infty}^{\infty} d\omega e^{-i\omega\tau}\varphi(\omega),
\label{solphih}
\end{eqnarray}
Eq.~(\ref{eqQ2A}) is solved as 
$
\tilde Q(\omega)=\lambda h(\omega)\varphi(\omega)
$
with 
$
h(\omega)={1/(-m\omega^2+m\Omega^2-i2m{\omega\gamma})}.
$
By inserting this solution (\ref{solQtau}) %with Eq.~(\ref{soltQA})
into the expression of $\phi_{\rm inh}(x)$, we have 
\begin{eqnarray}
\phi_{\rm inh}(x)=\lambda^2 \int d\tau\int {d\omega\over2\pi}e^{-i\omega\tau}
h(\omega)G_R(x-z(\tau))\varphi(\omega). 
\nonumber\\
\label{phiinh2}
\end{eqnarray}

%%%%%%%%%%%%%%%%%%%%%%%%%%%%%%%%%%%%%%%%%%%%%%%%%%%%%%%%%%%%%%%
%\section{Strong coupling $\Omega<\lambda^2/8\pi m$}
%%%%%%%%%%%%%%%%%%%%%%%%%%%%%%%%%%%%%%%%%%%%%%%%%%%%%%%%%%%%%%%
In the present paper, we consider the case $\Omega<\gamma$, 
in which the poles of $h(\omega)$ are located at
$
\omega=-i\Omega_\pm
$
where we defined
$
\Omega_{\rm \pm}
%={\lambda^2\over 8\pi m}
%\pm\sqrt{\biggl({\lambda^2\over 8\pi m}\biggr)^2-\Omega^2}. 
=\gamma\pm\sqrt{\gamma^2-\Omega^2}
$.

It is useful to verify that the detector is in thermal equilibrium at the Unruh temperature. 
The expectation value of energy of the harmonic oscillator is computed using the solution (\ref{solQtau}) 
with $h(\omega)$ as
\begin{eqnarray}
\langle E\rangle={m\over2}\left(\langle\dot Q^2(\tau)\rangle+\Omega^2\langle Q^2(\tau)\rangle
\right)={a\over 2\pi}
\end{eqnarray}
under the condition $\Omega_\pm\ll a$.  
Thus the law of the equipartition of energy with the Unruh temperature is satisfied
as a consequence of the Unruh effect. 

%%%%%%%%%%%%%%%%%%%%%%%%%%%%%%%%%%%%%%%%%%%%%%%%%%%%%%%%%%%%%%%%%%%
\section{Radiation from the Unruh-De Witt detector}
%%%%%%%%%%%%%%%%%%%%%%%%%%%%%%%%%%%%%%%%%%%%%%%%%%%%%%%%%%%%%%%%%%%
Since the detector is in the thermal equilibrium, one may expect that the would-be
radiation due to the thermal fluctuation is cancelled by the quantum interference effect.
Actually that is the case for the $1+1$ dimensional case.
The $1+3$ dimensional case has a similar structure of the cancellation, and 
 we misconcluded in Ref.~\cite{IYZ2013} 
 that the quantum radiation from the uniformly accelerating
Unruh-De Witt detector is completely cancelled. 
But more careful calculations show
that some part of the radiation  remains.
Our new conclusion is consistent with that in Ref.\cite{LH}, in which 
they also demonstrated  nonvanishing radiation flux. 
In the present paper, we give an analytic expression for the radiation
and some interpretation of the origin of the radiation.

\begin{figure}[t]
\begin{center}
    \includegraphics[width=7.5cm]{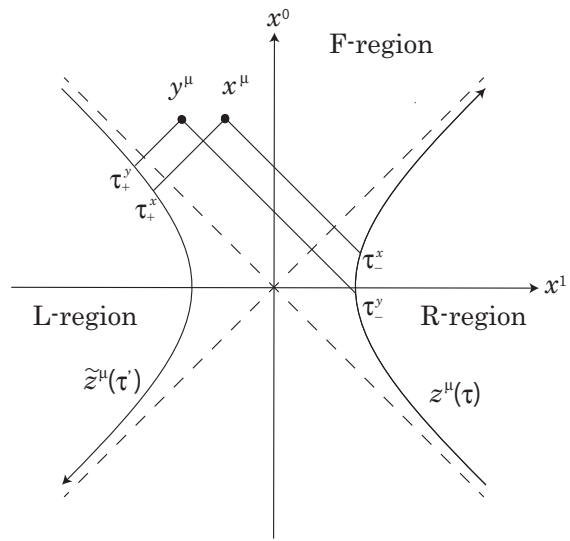}
%\hspace{1cm}
 %   \includegraphics[width=6.cm]{fig3b.eps}
\caption{
The R-region is defined by $x^1>|x^0|$, 
the L-region is $-x^1>|x^0|$, and the F-region 
is $x^0>|x^1|$. 
The hyperbolic curve $z^\mu(\tau)$ in the R-region is the trajectory 
of a uniformly accelerating Unruh-De Witt detector, 
while the hyperbolic curve in the L-region $\widetilde z^\mu(\tau)$ is the 
hypothetical trajectory obtained by an analytic continuation of the 
trajectory in the R-region. 
$\tau_-^x$ is defined by the proper time at which the detector's trajectory
intersects with the past lightcone of $x^\mu$.
On the other hand, for a point $y^\mu$ in the F-region, 
$\tau_+^y$ is defined by the proper time that the hypothetical 
detector's trajectory in the L-region intersects with the past 
lightcone of $y^\mu$. 
\label{fig:pinhphF}}
\end{center}
\end{figure}

In order to calculate the radiation from the detector, we evaluate the energy 
momentum tensor of the quantum field. 
First, we consider the two point function \cite{IYZ,IYZ2013}. 
Since the total radiation rate can be estimated from the flux in the F-region in Fig. \ref{fig:pinhphF}, 
we focus on the two point function,
\begin{eqnarray}
  &&
\langle\phi(x)\phi(y)\rangle-\langle\phi_{\rm h}(x)\phi_{\rm h}(y)\rangle
\nonumber\\
  &=&\langle\phi_{\rm inh}(x)\phi_{\rm h}(y)\rangle+\langle\phi_{\rm h}(x)\phi_{\rm inh}(y)\rangle+  \langle\phi_{\rm inh}(x)\phi_{\rm inh}(y)\rangle 
\nonumber\\
  &=&{-i\lambda^2\over (4\pi)^2\rho_0(x)\rho_0(y)}
  \int _{-\infty}^{+\infty}{d\omega\over2\pi}{e^{\pi\omega/a}\over e^{2\pi\omega/a}-1}
\nonumber
\\
&&\bigr[h(\omega)e^{-i\omega(\tau_-^x-\tau_+^y)}
-h(-\omega)e^{-i\omega(\tau_+^x-\tau_-^y)}\bigl],
\label{interferencew}
\end{eqnarray}
for $x,y\in$ F-region, where we defined $\rho_0(x)=a\sqrt{(-x_\mu x^\mu+1/a^2)^2/4+((x^0)^2-(x^1)^2)/a^2}$.
Here, $\tau_-^x$ is defined as the proper time at which the detector's trajectory
intersects with the past lightcone of a spacetime point $x$.
On the other hand, $\tau_+^x$ is the proper time at which the hypothetical detector's trajectory 
in the L-region intersects with the past lightcone of $x$ for $x\in$ F-region.
$\tau_\pm^y$ is defined in the same way. (See figure \ref{fig:pinhphF}). 

%%%%%%%%%%%%%%%%%%%%%%%%%%%%%%%%%%%%%%%%%%%%%%%%%%%%%%%%%%%%%%%%%
%%%%%%%%%%%%%%%%%%%%%%%%%%%%%%%%%%%%%%%%%%%%%%%%%%%%%%%%%%%%%%%%%

After performing the integration of (\ref{interferencew}),
the two point function symmetrized with respect to $x$ and $y$ is expressed as
\begin{eqnarray}
 &&[\langle\phi(x)\phi(y)\rangle-\langle\phi_{\rm h}(x)\phi_{\rm h}(y)\rangle]_S
\nonumber\\
&&=-{i\lambda^2\over(4\pi)^2\rho_0(x)\rho_0(y)}{1\over 2m}\left(I(x,y)+I(y,x)\right),
\label{phiphiphi}
\end{eqnarray}
where $I(x,y)$ is defined by
\begin{eqnarray}
&&I(x,y)=
\nonumber\\
&&-i\theta(\tau_-^y-\tau_+^x)\biggl[
{1\over \Omega_+\Omega_-}{a\over 2\pi}
+
{e^{-\Omega_-(\tau_-^y-\tau_+^x)}\over \Omega_--\Omega_+}{1\over \sin\pi\Omega_-/a}
\nonumber\\
&&
+{e^{-\Omega_+(\tau_-^y-\tau_+^x)}\over \Omega_+-\Omega_-}{1\over \sin\pi\Omega_+/a}
+\sum_{n=1}^\infty
      {(-1)^ne^{-na(\tau_-^y-\tau_+^x)}\over(\Omega_--na)(\Omega_+-na)}{a\over\pi}
  \biggr]
\nonumber\\
  && 
+i\theta(\tau_+^x-\tau_-^y)\biggl[
    {1\over \Omega_+\Omega_-}{a\over 2\pi}
    +\sum_{n=1}^\infty
    {(-1)^ne^{na(\tau_-^y-\tau_+^x)}\over(\Omega_-+na)(\Omega_++na)}{a\over\pi}
    \biggr].
\nonumber\\
\end{eqnarray}

\begin{figure}[t]
\begin{center}
    \includegraphics[width=5.cm]{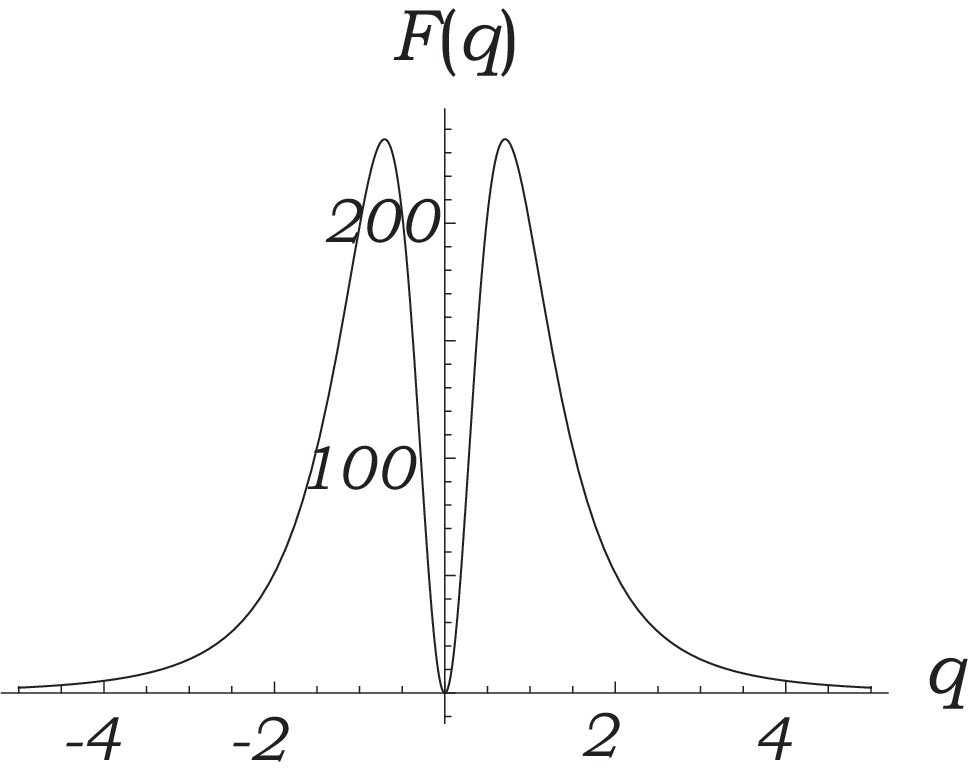}
\\
\vspace{8mm}
    \includegraphics[width=5.cm]{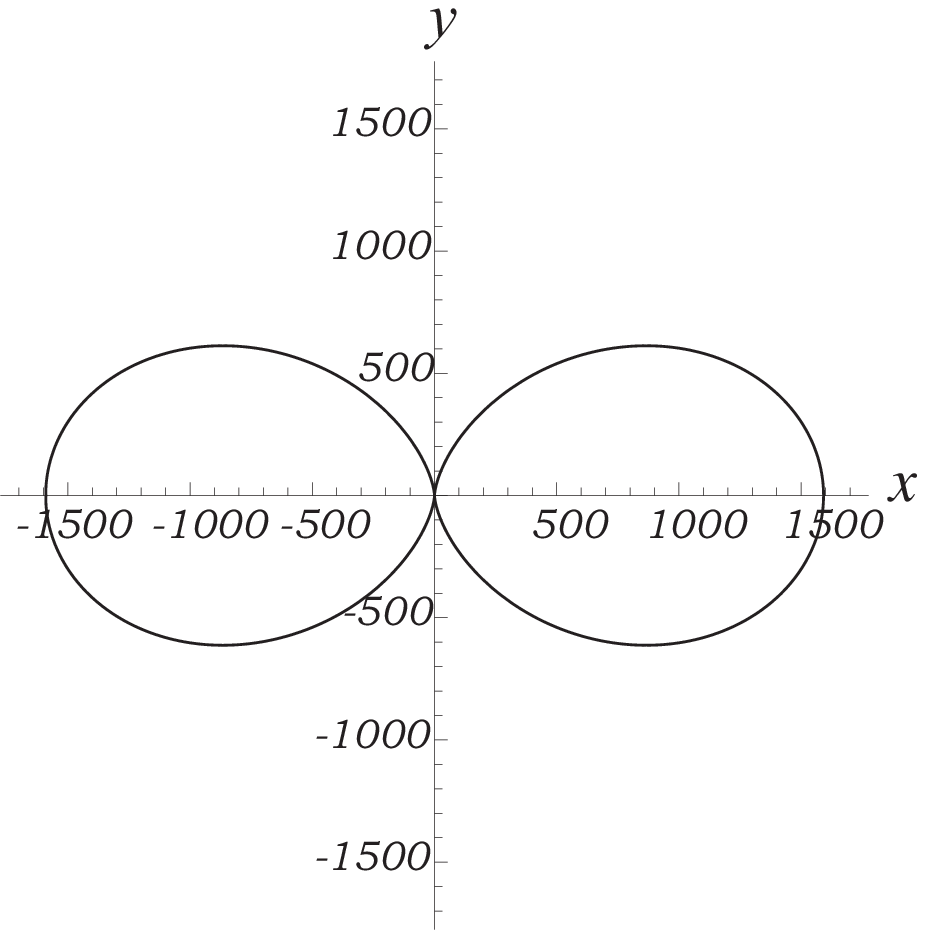}
\caption{Upper panel: ${\cal F}(q)$ as function of $q$, where we chose 
$\Omega/a=0.01$ and $\gamma/a=1$. 
Lower panel: Angular distribution 
of the flux $\sin^{-4}{\theta} {\cal F}(q(\tau_-,\theta))$ 
at $\tau_- = 0$, where we chose the same parameters as those of the upper panel.  
The coordinates $x$ and $y$ are $x^1$ and $\sqrt{(x^2)^2+(x^3)^2}$, respectively. 
\label{fig:funF}}
\end{center}
\end{figure}

We are now interested in the energy flux $f=-\sum_i T_{0i}n^i$, where $T_{0i}$ is the 
time and space component of the energy momentum tensor and $n^i$ is the 
unit vector $n^i=x^i/r$, which is computed from the two point function,
\begin{eqnarray}
\hspace{-4mm}
T_{0i}
=\lim_{y\rightarrow x}{\partial\over \partial x^0 }{\partial\over \partial y^i }
[\langle\phi(x)\phi(y)\rangle-\langle\phi_{\rm h}(x)\phi_{\rm h}(y)\rangle]_S.
\end{eqnarray}
Using the expression (\ref{phiphiphi}), we can derive an exact expression 
for the energy flux (cf. \cite{OYZ15, OYZ16}).   
The exact formula (see Appendix) is very complicated, but
in the case $\Omega<\gamma$, 
it can be very well approximated by
the following formula,
\begin{eqnarray}
f={a\lambda^2\over(4\pi)^2mr^2\sin^4\theta}{\cal F}(q,\Omega_+/a,\Omega_-/a),
\label{energyflux}
\end{eqnarray}
where we defined
\begin{eqnarray}
  &&{\cal F}(q,\Omega_+/a,\Omega_-/a)=
  {q^2\over(1+q^2)^3}\biggl[-\theta(q)\biggl\{{a^2\over\Omega_+\Omega_-}{1\over 2\pi}
\nonumber\\
&&
+{a\over {\Omega_-}-{\Omega_+}}
\biggl({{-q+\sqrt{1+q^2}}\over{q+\sqrt{1+q^2}}}\biggr)^{\Omega_-/a}
{1\over \sin\pi\Omega_-/a}\biggr\}
\nonumber\\  
    &&+\theta(-q)\biggl\{{a^2\over \Omega_+\Omega_-}{1\over 2\pi}\biggr\}\biggr]
\label{definitionofcalF}
\end{eqnarray}
and
%\begin{eqnarray}
%&&
$
q=a\left(t-r-{1/(2a^2 r)}\right)/\sin\theta.
$
%\end{eqnarray}
The upper panel of Fig.~\ref{fig:funF} exemplifies the function ${\cal F}(q)$ 
adopting ${\gamma/a}=1$ and ${\Omega/a}=0.01$.
The lower panel of Fig.~\ref{fig:funF} shows the corresponding angular plot of 
${\cal F}(q)/\sin^4\theta$ at $\tau_-=0$ (see also Refs.\cite{OYZ15,OYZ16})

The order of the energy radiation rate is roughly estimated as
\begin{eqnarray}
\hspace{-5mm}
  {dE\over dt}=\lim_{r\rightarrow\infty} r^2 \int d\Omega_{(2)} f
  \sim{a\lambda^2\over 4\pi m}{\cal F}\sim {a\lambda^2\over 4\pi m}
{a^2\over 2\pi \Omega^2}. 
\end{eqnarray}
This result is consistent with that of Ref.\cite{LH}, % in the equilibrium phase, 
though their result assumes the weak coupling case $\Omega>\gamma$.

%%%%%%%%%%%%%%%%%%%%%%%%%%%%%%%%%%%%%%%%%%%%%%%%%%%%%%%%%%%%%%%%%%
\section{Interpretation of the Result}
%%%%%%%%%%%%%%%%%%%%%%%%%%%%%%%%%%%%%%%%%%%%%%%%%%%%%%%%%%%%%%%%%%%

We will now point out that the physical origin of the remaining radiation is related to
the quantum entanglement of the vacuum between the left and the right Rindler wedges.
 Using the properties of the retarded Green function 
%and the advanced Green function, 
\begin{eqnarray}
&&\int d\tau G_R(x,z(\tau)) J(\tau)={J(\tau_-^x)\over 4\pi \rho_0(x)},
%\\
%&&\int d\tau G_A(x,\bar z(\tau)) f(\tau)={f(\tau_+^x)\over 4\pi \rho_0(x)}
\label{defGA}
\end{eqnarray}
for a function $J(\tau)$, the two point function (\ref{interferencew}) with $x,y\in$ 
F-region can be rewritten as 
\begin{eqnarray}
&&\langle \phi(x) \phi(y)\rangle  -\langle \phi_{\rm h}(x) \phi_{\rm h}(y) \rangle
=-i\lambda^2 \int \frac{d\omega}{2\pi} {e^{\pi\omega/a} \over e^{2\pi\omega/a}-1} 
\nonumber\\
&&~~
\int d\tau \int d\tau' e^{-i\omega(\tau-\tau')}
\biggl[G_R(x,z(\tau))G_{R}(y,\widetilde z(\tau'))h(\omega)
\nonumber\\
&&~~-G_{R}(x,\widetilde z(\tau))
G_R(y,z(\tau'))h(-\omega)\biggr],
\label{GGRRH}
\end{eqnarray}
where $\widetilde z(\tau)$ denotes the hypothetical trajectory 
in the L region. 
On the other hand, the correlation of the inhomogeneous term, 
which is cancelled by the interference term,  is given by \cite{IYZ,IYZ2013}, 
%%%%%%%%%%%%%%%%%%%%%%%%%%%%%%%%%%%%%%%%%%%%%%%%%%%%%%%%
\begin{eqnarray}
&&\langle \phi_{\rm inh}(x) \phi_{\rm inh}(y) \rangle
=-i\lambda^2 \int \frac{d\omega}{2\pi} {e^{2\pi\omega/a} \over e^{2\pi\omega/a}-1} 
\nonumber\\
&&~~
\int d\tau \int d\tau' e^{-i\omega(\tau-\tau')}
\biggl[G_R(x,z(\tau))G_{R}(y,z(\tau'))h(\omega)
\nonumber\\
&&~~-G_{R}(x,z(\tau))
G_R(y,z(\tau'))h(-\omega)\biggr].
\label{GGRR}
\end{eqnarray}
%%%%%%%%%%%%%%%%%%%%%%%%%%%%%%%%%%%%%%%%%%%%%%%%%%%%%%%%
These two correlations, (\ref{GGRRH}) and (\ref{GGRR}), look very similar but 
are different in the following two points, and both of them  
indicate that the remaining
two point function (\ref{GGRRH}) reflects the nonlocal correlation 
of the Minkowski vacuum state for the following two reasons. 

First, Eq.~(\ref{GGRR}) expresses the two point correlation of the field produced by the 
detector in the R-region, which is described by the retarded
Green function connecting two points on the trajectory $z^\mu(\tau)$ in the $R$-region 
(see Fig.~\ref{fig:pinhphF}). It is due to the fact that  the inhomogeneous part of the field $\phi_{\rm inh}$ 
is determined by the quantum fluctuations on the real trajectory (\ref{phiinh2}).
On the other hand, Eq.~(\ref{GGRRH}) is obtained by replacing one of the two points on the trajectory
$z^\mu(\tau)$ in the R-region with $\widetilde z^\mu(\tau)$ in the L-region.
This reflects the fact that the correlation function $\langle \phi_{\rm h} (x) \phi_{\rm inh} (y) \rangle$ contains the correlation between the R and the L regions. 
Namely, the entanglement of the quantum fluctuations between the R-region
and the L-region will be responsible for 
the remaining radiation  in Eq.~(\ref{GGRRH}). 

The second difference between (\ref{GGRRH}) and (\ref{GGRR}) is 
the numerical factors of $e^{\pi\omega/a}$
and $e^{2\pi\omega/a}$. 
%and the factor $e^{2\pi\omega/a}/(e^{2\pi\omega/a}-1)$ in Eqs.~(\ref{GGRRH}) 
%and (\ref{GGRR}) 
It is also a signature of the entanglement of  fields between the R-region and the L-region.
By introducing the Rindler coordinates in the R-region and the L-region,
the quantum field operator is constructed in each region, respectively, 
and we may write the field operator as \cite{UnruhWald,Higuchi},
\begin{eqnarray}
%$
\psi=\psi_{\rm R}\theta(x^1-x^0)+\psi_{\rm L}\theta(x^0-x^1),
%$
\label{appendixphi}
\end{eqnarray}
with 
\begin{eqnarray}
&&\psi_{\rm R}=\sum_j\left(u_j(x_{\tiny{\rm R}})\hat a_j+u_j^*(x_{\tiny{\rm R}})\hat a_j^\dagger\right),
\label{appendixphiR}
\\
&&\psi_{\rm L}=\sum_j\left(v_j(x_{\tiny{\rm L}})\hat b_j+v_j^*(x_{\tiny{\rm L}})\hat b_j^\dagger\right),
\label{appendixphiL}
\end{eqnarray}
where $\psi_{\rm R}$ and $\psi_{\rm L}$ are the quantum field operators,
$u_j(x_{\tiny{\rm R}})$ and $v_j(x_{\tiny{\rm L}})$ are the mode functions, 
and $\hat a_j (\hat a_j^\dagger)$ and $\hat b_j (\hat b_j^\dagger)$ are 
the annihilation (creation) operators of Rindler particles in the R-region and the L-region, respectively.
Accordingly the Rindler vacuum states, $|0,{\rm R}\rangle$ and $|0,{\rm L}\rangle$, 
are defined by  the annihilation operator, $\hat a_j$ or $\hat b_j$. 
The Minkowski vacuum state $|0,{\rm M}\rangle$ is expressed by the superposed
state of the excited states of the Rindler vacuum \cite{UnruhWald,Higuchi},
\begin{eqnarray}
|0,{\rm M}\rangle
=\prod_j\biggl[N_j\sum_{n_j=0}^\infty e^{-\pi n_j \omega_j/a}
|n_j,{\rm R}\rangle
\otimes|n_j,{\rm L}\rangle\biggr],
\label{Minkowskivac}
\nonumber\\
\end{eqnarray}
where $|n_j,{\rm R}\rangle$ and $|n_j,{\rm L}\rangle$ are the
$n$th excited states of the mode $j$ for 
the Rindler particles in the R-region and in the L-region, respectively. 
$\omega_j$ is the energy of a Rindler particle of the mode $j$,
and $N_j=\sqrt{1-e^{-2\pi \omega_j/a}}$. 
This expression describes the entanglement of the Minkowski vacuum state, 
as the entangled states of the R-region and the L-region.

Let us consider the field operator of the form (\ref{appendixphiR}) 
but with  $u_j(x_R)$ being replaced by another  function $\tilde{u}_j(x_R)$, 
which we define 
$\widetilde \psi(x_{\rm R})=\sum_j\left(\tilde u_j(x_{\tiny{\rm R}})\hat a_j+\tilde u_j^*(x_{\tiny{\rm R}})\hat a_j^\dagger\right)$. 
% as a simplified model of $\psi_{\rm inh}(x_R)$.
By choosing points, $x$ and $y$ in the R-region and in the L-region, respectively,
the correlation function $\langle 0,{\rm M}|\widetilde\psi(x)\psi(y)|0,{\rm M}\rangle$ can be obtained as
\begin{eqnarray}
&&\langle 0,{\rm M}| \widetilde\psi(x_R)\psi_{}(y_L) |0,{\rm M}\rangle
=\sum_j(\tilde{u}_j(x_{\tiny{\rm R}})v_j(y_{\tiny{\rm L}})
\nonumber\\
&&~
+\tilde{u}_j^*(x_{\tiny{\rm R}})v_j^*(y_{\tiny{\rm L}})){e^{\pi\omega_j/a}\over e^{2\pi\omega_j/a}-1}. 
\label{MVtwopoint}
\end{eqnarray}
Here the factor $e^{\pi\omega/a}/( e^{2\pi\omega/a}-1)$ appears when 
the two points are chosen  in the R-region and the L-region. 
This comes from the relations $\langle 0,{\rm M}|\hat b_j\hat a_j|0,{\rm M}\rangle
=\langle 0,{\rm M}|\hat a_j^\dagger\hat b_j^\dagger|0,{\rm M}\rangle\propto e^{\pi\omega/a}/(e^{2\pi\omega/a}-1)$,
and the same factor appears in (\ref{GGRRH}).

On the other hand, when two points $x$ and $y$ are in the R-region, the two point correlation 
function becomes
\begin{eqnarray}
&&\langle 0,{\rm M}| \widetilde\psi(x_R) \psi_{}(y_R) |0,{\rm M}\rangle
=\sum_j\bigl(\tilde{u}_j(x_{\tiny{\rm R}}){u}_j^*(y_{\tiny{\rm R}}){e^{2\pi\omega_j/a}\over e^{2\pi\omega_j/a}-1}
\nonumber\\
&&~
+\tilde{u}_j^*(x_{\tiny{\rm R}}){u}_j(y_{\tiny{\rm R}}){1\over e^{2\pi\omega_j/a}-1}\bigr).
\label{MVtwopointRR}
\end{eqnarray}
Note that a different numerical  factor $e^{2\pi\omega/a}$ appears in the numerator.
This comes from the relations $\langle 0,{\rm M}|\hat a_j\hat a_j^\dagger|0,{\rm M}\rangle
\propto e^{2\pi\omega/a}/(e^{2\pi\omega/a}-1)$ and 
 $\langle 0,{\rm M}|\hat a_j^\dagger\hat a_j|0,{\rm M}\rangle\propto 1/(e^{2\pi\omega/a}-1)$,
 and this is nothing but the numerical factor in  (\ref{GGRR}). 
By changing the integration variable from $\omega$ to  $\omega'=-\omega$, 
the function $1/(e^{2\pi\omega/a}-1)$ is expressed as $e^{2\pi\omega'/a}/(1-e^{2\pi\omega'/a})$
and becomes the same numerical factor.

Thus, the above arguments show that the difference in the numerical factors of $e^{\pi\omega/a}$
and $e^{2\pi\omega/a}$ can be interpreted as an indication of the entanglement 
of the Minkowski vacuum between the right Rindler wedge and the left Rindler wedge as Eq.~(\ref{Minkowskivac}). 
%%%%%%%%%%%%%%%%%%%%%%%%%%%%%%%%%%%%%%%%%%%%%%%%%%%%%%%%%%%%%%%%%%
\section{Summary and Conclusions}
%%%%%%%%%%%%%%%%%%%%%%%%%%%%%%%%%%%%%%%%%%%%%%%%%%%%%%%%%%%%%%%%%%%
In summary the influence of the detector in the quantum vacuum is generated 
in the R-region, which is described by $\phi_{\rm inh}(x)$, and propagates into the F-region.
However, the system  cannot be closed within the R-region. 
As we showed,  
the remaining energy flux in the F-region, which can be calculated from the two point functions there, 
depends on the interference  between  $\phi_{\rm inh}(x)$ and $\phi_{h}(x)$ in the F-region. 
Due to the causality, properties of the quantum field  $\phi_{\rm h}(x)$ in the F-region are influenced
by the properties of the quantum states   not only in  the R-region but also in the L-region. 
Since the Minkowski vacuum is entangled between these two regions, 
the correlation function of  $\phi_{\rm inh}(x)$ and $\phi_{h}(x)$ contains the information of the entanglement
of the Minkowski vacuum.
If there was no entanglement, the energy flux would be completely cancelled out and vanish.
Thus, we can conclude that the  remaining radiation 
is a consequence of the nonlocal correlation (or the entanglement) of the Minkowski vacuum between the R and L regions,
and it may be called the quantum radiation.

Detectability of the quantum radiation is an interesting issue, and 
in order to discuss it, we first need to extend the present calculation
to more realistic systems.
It is also necessary to satisfy the condition that 
thermalization time
(or the relaxation time)  $\tau_R={8\pi m/ \lambda^2}=\gamma^{-1}$ \cite{LH}, 
with which the system becomes in an equilibrium phase, must be shorter 
than the time during which a uniform acceleration is maintained. 
We hope to discuss these issues in future publications.

%%%%%%%%%%%%%%%%%%%%%%%%%%%%%%%%%%%%%%%%%%%%%%%%%%%%%%%%
\def\aM{a_{ \rm M}}
\def\aR{a_{ \rm R}}
\def\aL{a_{ \rm L}}
\def\phiL{\phi_{ \rm L\omega}}
\def\phiR{\phi_{ \rm R\omega}}
\def\FL{F_{ \rm L\omega}}
\def\FR{F_{ \rm R\omega}}
%%%%%%%%%%%%%%%%%%%%%%%%%%%%%%%%%%%%%%%%%%%%%%%%%%%%%%%%
\section*{Acknowledgments}
This work was supported by MEXT/JSPS KAKENHI Grant Number 15H05895,
and the Grant-in-Aid for Scientific research from the Ministry of Education, Science, Sports, and Culture, Japan, Nos.
23540329. 

%\end{multicols}
%\begin{multicols}{1}
%\onecolumn

\begin{widetext}

\section*{Appendix}
In the appendix we just show the result of 
the exact formula for the energy flux (\ref{energyflux}) with
\begin{eqnarray}
  &&\hspace{0cm}{\cal F}(q,\widetilde\Omega_+,\widetilde\Omega_-)={q^2\over(1+q^2)^3}\biggl[-\theta(q)\biggl\{{1\over\widetilde\Omega_+}{1\over\widetilde\Omega_-}{1\over 2\pi}
+{\xi^{\widetilde\Omega_-}(q)\over {\widetilde\Omega_-}-{\widetilde\Omega_+}}
{1\over \sin\pi\widetilde\Omega_-}+{\xi^{\widetilde\Omega_+}(q)\over {\widetilde\Omega_+}-{\widetilde\Omega_-}}
{1\over \sin\pi\widetilde\Omega_+}
 \nonumber
\\
&&\hspace{0cm} +{\xi(q)\over {\widetilde\Omega_-}-{\widetilde\Omega_+}}
\biggl({1\over{1-{\widetilde\Omega_+}}}{}_2F_1(1,1-\widetilde\Omega_+,2-\widetilde\Omega_+;-\xi(q))   
-{1\over{1-\widetilde\Omega_-}}{}_2F_1(1,1-\widetilde\Omega_-,2-\widetilde\Omega_-;-\xi(q))\biggr){1\over\pi}\biggr\}
\nonumber\\
&&
+\theta(-q)\biggl\{{1\over \widetilde\Omega_+}{1\over\widetilde\Omega_-}{1\over 2\pi}
+{\xi^{-1}(q)\over \widetilde\Omega_--\widetilde\Omega_+}
\biggl(
{1\over{1+\widetilde\Omega_-}}{}_2F_1(1,1+\widetilde\Omega_-,2+\widetilde\Omega_-;-\xi^{-1}(q))   
\nonumber\\
  &&\hspace{0cm}  -{1\over{1+\widetilde\Omega_+}}{}_2F_1(1,1+\widetilde\Omega_+,2+\widetilde\Omega_+;-\xi^{-1}(q))\biggr){1\over\pi}\biggr\}\biggr]
 -2{q\over(1+q^2)^{5/2}}
  \biggl[-\theta(q)\biggl\{-{{\widetilde\Omega_-}\over {\widetilde\Omega_--\widetilde\Omega_+}}
{\xi^{\widetilde\Omega_-}(q)\over \sin\pi\widetilde\Omega_-} 
\nonumber
\\&&
-{\widetilde\Omega_+\over \widetilde\Omega_+-\widetilde\Omega_-}
{\xi^{\widetilde\Omega_+}(q)\over \sin\pi\widetilde\Omega_+} 
+{\xi(q)\over{\widetilde\Omega_--\widetilde\Omega_+}}
\biggl(-{1\over{1-\widetilde\Omega_+}}
{}_2F_1(2,1-\widetilde\Omega_+,2-\widetilde\Omega_+;-\xi(q))  
+{1\over{1-\widetilde\Omega_-}}
\nonumber\\
&&
{}_2F_1(2,1-\widetilde\Omega_-,2-\widetilde\Omega_-;-\xi(q))\biggr){1\over\pi}\biggr\}     
+\theta(-q)\biggl\{
{\xi^{-1}(q)\over{\widetilde\Omega_--\widetilde\Omega_+}}
\biggl(
-{1\over{1+\widetilde\Omega_+}}{}_2F_1(2,1+\widetilde\Omega_+,2+\widetilde\Omega_+;-\xi^{-1}(q)) 
\nonumber\\
  &&\hspace{0cm}
+{1\over{1+\widetilde\Omega_-}}{}_2F_1(2,1+\widetilde\Omega_-,2+\widetilde\Omega_-;-\xi^{-1}(q))\biggr)
{1\over\pi} \biggr\}\biggr] 
-{1\over(1+q^2)^2}\biggl[-\theta(q)\biggl\{
  -{\widetilde\Omega_-^2\over \widetilde\Omega_--\widetilde\Omega_+}{\xi^{\widetilde\Omega_-}(q)\over \sin\pi\widetilde\Omega_-}
-{\widetilde\Omega_+^2\over \widetilde\Omega_+-\widetilde\Omega_-}
\nonumber\\  
&& \hspace{0cm}
{\xi^{\widetilde\Omega_+}(q)\over \sin\pi\widetilde\Omega_+}  
+{1\over \widetilde\Omega_--\widetilde\Omega_+}\biggl({\xi(q)\over 1-\widetilde\Omega_-}{}_2F_1(2,1-\widetilde\Omega_-,2-\widetilde\Omega_-;-\xi(q))  
-{\xi(q)\over 1-\widetilde\Omega_+}
{}_2F_1(2,1-\widetilde\Omega_+,2-\widetilde\Omega_+;-\xi(q))   
\nonumber\\     
   &&\hspace{0cm}   
-{2\xi^2(q)\over 2-\widetilde\Omega_-}
{}_2F_1(3,2-\widetilde\Omega_-,3-\widetilde\Omega_-;-\xi(q))   
+{2\xi^2(q)\over 2-\widetilde\Omega_+}
{}_2
F_1(3,2-\widetilde\Omega_+,3-\widetilde\Omega_+;-\xi(q))\biggr){1\over\pi}\biggr\} 
+\theta(-q)\biggl\{
\nonumber\\     
   &&\hspace{0cm}   
{1\over \widetilde\Omega_--\widetilde\Omega_+}\biggl(-{\xi^{-1}(q)\over 1+\widetilde\Omega_-}
{}_2F_1(2,1+\widetilde\Omega_-,2+\widetilde\Omega_-;-\xi^{-1}(q)) 
+{\xi^{-1}(q)\over 1+\widetilde\Omega_+}
{}_2F_1(2,1+\widetilde\Omega_+,2+\widetilde\Omega_+;-\xi^{-1}(q)
\nonumber\\     
   &&\hspace{0cm} 
+{2\xi^{-2}(q)\over 2+\widetilde\Omega_-}{}_2F_1(3,2+\widetilde\Omega_-,3+\widetilde\Omega_-;-\xi^{-1}(q))  
   -{2\xi^{-2}(q)\over 2+\widetilde\Omega_+}
{}_2F_1(3,2+\widetilde\Omega_+,3+\widetilde\Omega_+;-\xi^{-1}(q))\biggr){1\over\pi}\biggr\}\biggr],
 \end{eqnarray}
where we defined
$ \xi(q)=({-q+\sqrt{1+q^2})/ (q+\sqrt{1+q^2}})$, 
$\widetilde\Omega_+=\Omega_+ / a $ 
and
 $\widetilde\Omega_-=\Omega_- / a$ .
This complicated formula is well approximated by  Eq.~(\ref{definitionofcalF}) in the case 
$\Omega<\gamma$.

%\end{multicols}
%\begin{multicols}{2}
\end{widetext}

%\end{multicols}
\end{document}